\documentstyle[12pt]{article}

\newcommand{\be}{\begin{equation}}
\newcommand{\ee}{\end{equation}}
\newcommand{\ba}{\begin{eqnarray}}
\newcommand{\ea}{\end{eqnarray}}
\newcommand{\sect}[1]{\begin{center}{\Large #1}\end{center}}

\newcommand{\halpha}{\hat{\alpha}}
\newcommand{\hbeta}{\hat{\beta}}

\begin{document}
\thispagestyle{empty}
\noindent
\begin{center}
\begin{flushright}
{IP/BBSR/94-32}
\end{flushright}
\vspace*{30mm}
{\sc Duality Invariant Superstring Actions}\\
\vspace*{20mm} \large Abbas Ali \\
\vspace*{10mm} \normalsize {\it Institute of Physics,
Bhubaneswar-751005, India\\}
\vspace*{20mm}
\large Abstract \\
\end{center}
Worldsheet supersymmetric string action is written in duality
invariant form for flat as well as curved backgrounds. First the
action in flat backgrounds is written by introducing auxiliary fields.
We also give the superfield form of these actions and obtain the
off-shell supersymmetry algebra. The action has a modified
Lorentz invariance and supersymmetry and reduces to the
usual form when the auxiliary fields are eliminated using their
equations of motion. Supersymmetric nonlinear sigma model in curved
backgrounds
is also written in manifestly duality invariant form when
the background metric and tortion fields are independent of
some of the coordinates.

\newpage
 \sect{1. Introduction}
Symmetries of string theory play an important role in unravelling
its basic structure. One such symmetry, namely target space duality,
has received a great deal of attention recently \cite{d1}-\cite{alok}.
It is a generalization of $R \rightarrow 1/R$ duality
and interchanges the winding mode and Kluza-Klein excitations
(see \cite{gpr} for a comprehensive review.) In its usual form,
at least for the curved backgrounds, target space duality has
been formulated as a symmetry of the string effective
action only\cite{mv}.  To see whether it is a symmetry to all
orders in $\alpha^\prime$, it is useful to study
this duality from worldsheet point of view. Recently an
off-shell duality invariant formulation of worldsheet action was
given in Ref.\cite{tseytlin} by introducing extra `dual'
coordinate \cite{gpr}. In this case local Lorentz invariance is
not manifest. The action has a  symmetry which reduces to the
usual local Lorentz symmetry if the dual coordinates are
integrated out using their equations of motion. This is possible
since the dual coordinates are auxiliary fields on the
worldsheet. Auxiliary nature of the dual coordinates is also
evident from the fact that the Weyl anomaly is absent in $D=26$.
It has also been shown that one has to introduce the dual
coordinates to make the duality invariance manifest even for the
worldsheet equations of motion\cite{eom}.

Similar developments have taken place in the case of closed NSR
strings as well. Duality transformations of the worldsheet equations of
motions were discussed in \cite{siegel}. It was shown in \cite{dasjmu}
that, in superspace, the first order formalism adopted to write the
bosonic string equations of motion can be extended to the fermionic case
to implement the duality  transformations.

Recently a worldsheet duality invariant action for the bosonic
sector of the toroidally compactified four dimensional heterotic
string in curved backgrounds was written by Schwarz and
Sen\cite{schsen}. Thus, in absence of anomalies, the toroidally
compactified heterotic string theory should have the duality
symmetry order by order in Newton's constant, provided that at
each order the full nonperturbative $\alpha^\prime$ dependence
is taken into account.

In this paper we address the case of superstrings both in flat
backgrounds and nonlinear sigma model in curved gravitational
and antisymmetric tensor field backgrounds when the backgrounds
are independent of some of the coordinates.

The paper is organized in the following way.
Section 2 reviews the requisite parts of Ref.\cite{schsen}.
Section 3 starts by fixing the notation for studying
supersymmetric action in flat backgrounds. A duality invariant
worldsheet supersymmetric action is obtained by defining a
fermionic `vector' field under duality transformations.
Supersymmetric transformations are written in manifestly
duality invariant form. Both these symmetries are modified in the
compactified sector though they are equivalent to the usual
symmetries on shell. Supercurrent is written in duality
invariant form. Duality invariant supersymmetry is shown to be
off-shell closed even in the absence of usual auxiliary fields of the
supermultiplet. Therefore  we can introduce our superfields as
chiral multiplets.  In Section 4 we start with a nonlinear
supersymmetric sigma model in curved backgrounds when
the backgrounds are independent of
some of the coordinates corresponding to the compact target
coordinates of string theory. A duality invariant form is
obtained for this action.
Supersymmetric transformations are written in duality invariant
form. We end with some discussions and conclusions in Section 5.
Some definitions are collected in Appendix A.

\sect{2. Review of Duality Invariant Action \\
for Bosonic Strings}

A worldsheet action for the bosonic sector of the compactified
heterotic string in curved backgrounds was written down by
Schwarz and Sen \cite{schsen}. In this paper,
since we ignore the gauge fields, we are
concerned with bosonic or fermionic strings only.
The appropriately truncated action has
the form
\ba
S_B=-\frac{1}{2\pi}\int
d^2\sigma\{&\!g_{\mu\nu}\eta^{\halpha\hbeta}\partial_{\halpha}
X^\mu\partial_{\hbeta}X^\nu -
b_{\mu\nu}\epsilon^{\halpha\hbeta}\partial_{\halpha}
X^\mu\partial_{\hbeta}X^\nu \nonumber\\
&-\partial_0y^aL_{ab}\partial_1y^b-\partial_1y^a
(LML)_{ab}\partial_1y^b \;\;\;\},\label{baction}
\ea
where the uncompactified coordinates have Greek indices $\mu,
\nu, \cdots$ and Greek indices with a caret $\halpha, \hbeta,
\cdots $ are the worldsheet indices. Roman indices $a, b, \cdots $
transform vectorially under duality transformations. The coordinates
$y^a$ have the internal coordinates $y^i$ and its dual ${\tilde y}_i$
as its components where the indices $i, j, \cdots $ label internal
(compactified) coordinates. The $y^a$ equation of motion which follows
from this action can be written in the following form
\be
\partial_0y^a=-{(ML)^a}_b\partial_1y^b.\label{beom}\\
\ee
Here $\partial_0$ and $\partial_1$ are the worldsheet time and
space derivatives respectively. $L$ is a matrix which remains
invariant under the $O(d, d)$ transformations $\Omega$, $i.e.$
$L \rightarrow \Omega^TL\Omega = L$. The matrix $M$ transforms
as $M\rightarrow \Omega M\Omega^T$. Explicit forms of $L$ and $M$ are,
\be
L\;= \;\left(\matrix{{0}&{I_d}\cr {I_d}&{0}}\right),
M\;=\;\left(\matrix{{G^{-1}}&{-G^{-1}B}\cr
{BG^{-1}}&{G-BG^{-1}B}}\right), \label{landm}
\ee
where $G_{ij}$ and $B_{ij}$ are the internal parts of the metric
and antisymmetric tensor backgrounds. We are taking the total
metric and antisymmetric tensor fields to be in block diagonal
form in the target space and internal coordinates:
\be
g_{AB}\;=\;\left(\matrix{{g_{\mu\nu}}&{0}\cr
{0}&{G_{ij}}}\right);\;\;\;
b_{AB}\;=\;\left(\matrix{{b_{\mu\nu}}&{0}\cr
{0}&{B_{ij}}}\right) \label{block}
\ee
The equation of motion (\ref{beom}) has the component form
\ba
&\partial_0y^i=G^{ij}B_{jk}\partial_1y^k-
G^{ij}{\partial_1}{\tilde y}_j,&{}\label{tyeom}\\
&\partial_0{\tilde y}_i= B_{ij}G^{jk}B_{kl}\partial_1y^l
-G_{ij}\partial_1y^j- B_{ij}G^{jk}{\partial_1}{\tilde y}_k
&{}\label{yeom}
\ea
where (\ref{tyeom}) and (\ref{yeom}) are the ${\tilde y}_i$ and
$y^i$ equations of motion respectively.

The modified reparametrization under which the internal part of
(\ref{baction}) is invariant has the form
\be
\delta y^a=\xi^1\partial_1y^a-\xi^0(ML)^a_by^b.\label{blorentz}
\ee
The symmetric and traceless energy-momentum tensor for
(\ref{baction}) can be found either by Noether method or by
varying its generalization obtained by coupling it to the worldsheet
metric $h_{\halpha\hbeta}$ which gives Weyl and reparametrization
invariant action. Such action was also written in
\cite{schsen}. By varying it with respect to the metric
$h_{\halpha\hbeta}$ we get the following expressions for the
compactified sector
\ba
T_{00}=-T_{11}=
&-\frac{1}{2}\partial_1y^a(LML)_{ab}\partial_1y^b,\nonumber\\
T_{01}=T_{10}=&-\frac{1}{2}\partial_1y^aL_{ab}\partial_1y^b.
\label{bemtensor}
\ea
Using (\ref{yeom}) to eliminate ${\tilde y}_i$, we find that the
eqs.(\ref{bemtensor}) reduce to the usual form of the energy momentum
tensor.

\sect{3. Superstrings in Flat Backgrounds}
In this section we give a manifestly duality invariant
worldsheet supersymmetric string action in flat backgrounds and
discuss some of its aspects. The two component Majorana-Weyl
fermions in the uncompactified sector are denoted by $\psi^\mu$
and the ones in the compactified sector by $\psi^i$.
We use the following representation of $\gamma$-matrices in two
dimensions.
\be
\gamma^0 =\sigma^2;\;\;\gamma^1=i\sigma^1;\;\;
\gamma^5=\gamma^0\gamma^1=\sigma^3.\label{gammas}
\ee

In this
section we restrict ourself to flat backgrounds and also put
the antisymmetric tensor field $B_{ij}$ equal to zero. In this case
$M$ becomes $M=$diag$(G^{ij}, G_{ij}).$
The fermionic action which gives the supersymmetric action
when combined with the bosonic counterpart (whose duality
invariant form is given by (\ref{baction}) is
\be
S_F=\frac{i}{2\pi}\int d^2\sigma
(G_{ij}\bar{\psi}^i\gamma^{\halpha} \partial_{\halpha}\psi^j
+G_{\mu\nu}\bar{\psi}^\mu\gamma^{\halpha}\partial_{\halpha}\psi^\nu).
\label{fermion}
\ee
We have to put this action in duality invariant form
such that, when combined with $S_B$ given by eqn.(\ref{baction}),
it gives the desired duality invariant superstring action.
To write this action in duality invariant form we have to complete
the fermionic dual multiplet
\be
\psi^a \equiv \left(\matrix{{\psi^i}\cr
{\tilde{\psi_i}}}\right).
\ee

We now show that the fermionic action
is duality invariant without introducing new fields. This is due
to the fact that the fermionic action is already of the first order
form. Similar results were used in
Ref.\cite{schsen1} to write the supersymmetric version of the
duality invariant Maxwell action. To implement it in the
present case, we start with the usual supersymmetry transfomations
\be
\delta_SX^\mu =\bar{\epsilon}\chi^\mu\;;
\delta_S\chi^\mu=-i\gamma^{\halpha}\partial^{\halpha}X^\mu
\epsilon,\label{noncompsusy}\\
\ee
\be
\delta_Sy^i=\bar{\epsilon}\psi^i\;;
\delta_S\psi^i=-i\gamma^{\halpha}\partial^{\halpha}y^i\epsilon.\\
\label{compsusy}
\ee
The transformations in eq.(\ref{noncompsusy}) remain unchanged
under duality and we have to concentrate on eq.(\ref{compsusy}).
We generalize the first transformation in (\ref{compsusy}) as
\be
\delta_Sy^a=\bar{\epsilon}\psi^a.
\ee
To generalize the second transformation of eq.(\ref{compsusy}) we
write
\be
\delta_S\psi^a = -i\gamma^1\partial_1y^a\epsilon +
i{(ML)^a}_b\gamma^0 \partial_1y^b\epsilon.\label{compsusy1}
\ee
This equation has two components
\be
\delta _S\psi^i =
-i\gamma^1\partial_1y^i\epsilon + i\gamma^0
G^{ij}\partial_1{\tilde y}_j\label{nsusy0}
\ee
and
\be
\delta _S{\tilde \psi}_i =
-i\gamma^1\partial_1{\tilde y}_i\epsilon + i\gamma^0
G_{ij}\partial_1y^j.\label{nsusy1}
\ee
Eqn.(\ref{nsusy0}) reduces to (\ref{compsusy}) by
using equation of motion (\ref{beom}). We also find that the choice
\be
{\tilde\psi}_i = G_{ij}\gamma^5\psi^j.\label{tildepsi}
\ee
is consistent with eq.(\ref{compsusy1}).
This explicit form for ${\tilde\psi}_i$ and above duality
invariant form of the susy transformation can now be used
to cast the fermionic action in appropriate form. From
eqn.(\ref{tildepsi})
\be
\bar{\tilde\psi}_i = -G_{ij}\bar{\psi}^j\gamma^5.\label{tildepsi1}
\ee

We now write (\ref{fermion}) using (\ref{tildepsi}) or
(\ref{tildepsi1}) as
\be
S_F=\frac{i}{4\pi}\int d^2\sigma({\bar\psi}^i
\gamma^{\halpha} \gamma^5\partial_{\halpha}{\tilde\psi}_i
+ {\bar{\tilde\psi}}_i\gamma^{\halpha}
\gamma^5\partial_{\halpha}\psi^i
+2G_{\mu\nu}\bar{\psi}^\mu\gamma^{\halpha}
\partial_{\halpha}\psi^\nu).
\ee
Here we have also used the fact that the gravitational background is
flat. Using the  matrix $L$ from eqn.(\ref{landm}) we can write
this action in the manifestly duality invariant form as
\be
S_{F1}=\frac{i}{2\pi}\int d^2\sigma (\frac{1}{2}L_{ab}{\bar\psi}^a
\gamma^{\halpha} \gamma^5\partial^{\halpha}\psi^b
+G_{\mu\nu}\bar{\psi}^\mu\gamma^{\halpha}
\partial_{\halpha}\psi^\nu).\label{dfaction}
\ee
One could also write, using (\ref{tildepsi}) and (\ref{tildepsi1}),
\be
S_F=\frac{i}{2\pi}\int d^2\sigma
(G^{ij}\bar{{\tilde\psi}}_i\gamma^{\halpha}
\partial_{\halpha}{\tilde\psi}_j
+G_{\mu\nu}\bar{\psi}^\mu\gamma^{\halpha}
\partial_{\halpha}\psi^\nu)
\label{fermion1}
\ee
which, when added with (\ref{fermion}), has the duality invariant form
\be
S_{F2}=\frac{i}{2\pi}\int d^2\sigma
(\frac{1}{2}(LML)_{ab}{\bar\psi}^a
\gamma^{\halpha}\partial_{\halpha}\psi^b
+G_{\mu\nu}\bar{\psi}^\mu\gamma^{\halpha}
\partial_{\halpha}\psi^\nu).\label{dfaction1}
\ee
One notices the both of the forms, eqs.(\ref{dfaction}) and
(\ref{dfaction1}), of the fermionic action have usual Lorentz
invariance of the original action.

We can also write the supercurrent,
\ba
J_{\halpha}& =\frac{1}{2}\gamma^{\hbeta}\gamma_{\halpha}
\psi^A\partial_{\hbeta}X^BG_{AB}\nonumber\\
 & =\frac{1}{2}\gamma^{\hbeta}\gamma_{\halpha}
\psi^\mu\partial_{\hbeta}X^\nu g_{\mu\nu}
+\frac{1}{2}\gamma^{\hbeta}\gamma_{\halpha}
\psi^i\partial_{\hbeta}y^jG_{ij},
\ea
in the duality invariant form (upto equations of motion) as,
\be
J_{\halpha}=
\frac{1}{2}\gamma^{\hbeta}\gamma_{\halpha}
\psi^\mu\partial_{\beta}X^\nu g_{\mu\nu}
+\frac{1}{2}\gamma^1\gamma_{\halpha}
\psi^a(LML)_{ab}\partial_1y^b.\\
\ee

Next we give a superfield formulation of the action.
Since there is no explicit reparametrization invariance in the
bosonic sector, it is absent in the superfield form as well.
With the help of the following superderivatives,
\be
D_0=\frac{\partial}{\partial\bar{\theta}}-i\gamma^0\theta
\partial_0;
D_1=\frac{\partial}{\partial\bar{\theta}}-i\gamma^1\theta
\partial_1;
\ee
and the chiral superfield
\be
Y^a =y^a +i\bar{\theta}\psi^a
\ee
an appropriate superfield action can be written in the following
form:
\be
S_F=-\frac{i}{4\pi}\int d^2\sigma d^2\theta
[\bar{D}_1Y^a(LML)_{ab}D_1Y^b + L_{ab}\bar{D}_0Y^a\gamma^5D_1Y^b].
\ee
Notice that the superfield in this case is chiral, $i.e$, it does
not contain the auxiliary field of the supersymmetry multiplet.
The supersymmetry algebra closes off-shell and we have
\be
\{\delta_{S1}, \delta_{S2}\}y^a =
2i\bar{\epsilon}_1\gamma^1\epsilon_2\partial_1 y^a -
2i{(ML)^a}_b\bar{\epsilon}_1\gamma^0\epsilon_2\partial_1y^b.
\ee
\sect{4. Nonlinear $\sigma$-Models in Curved Backgrounds}
In this section we start with a supersymmetric nonlinear sigma model
in gravitational and antisymmetric tensor field backgrounds which are
independent of some of the coordinates and discuss how to write down a
manifestly duality invariant form of the action. We take the action
given in Ref.\cite{ss1,ss2}
\ba
S_{NSM}=-\frac{1}{2\pi}\int
d^2\sigma [\!&g_{AB}\partial_{\halpha}X^A\partial^{\halpha}X^B -
ig_{AB}{\bar\psi}^A(\slash\!\!\!\!{\cal D}\psi)^B\nonumber\\
&+b_{AB}\epsilon^{\halpha\beta}
\partial_{\halpha}X^A\partial_{\hbeta}X^B
 -\frac{i}{2}\partial_{\halpha}(b_{AB}
{\bar\psi}^A\gamma^5\gamma^{\halpha}\psi^B)\nonumber\\
& + \frac{1}{8}{\cal R}_{ABCD}
{\bar\psi}^A(1+\gamma^5)\psi^C{\bar\psi}^B(1+\gamma^5)\psi^D\;]
\label{ss1}
\ea
where
\be
({\cal D}_{\halpha}\psi)^A \equiv \partial_{\halpha}\psi^A +
(\eta_{\halpha\hbeta}\Gamma^A_{BC} -
\epsilon_{\halpha\hbeta}{S^A}_{BC})\partial^{\hbeta}X^B\psi^C,
\ee
and ${\cal R}_{ABCD}$, $\Gamma_{BC}^A$ and $S^A_{BC}$ are defined
in Appendix A.

In the case when the backgrounds
are independent of some of the coordinates $y^i$'s and depend
only on the rest of the coordinates $X^\mu$'s, the
background metric $g_{AB}$ and the antisymmetric tensor field
can be written in a block diagonal form (\ref{block}).
Also $S_{ABC}$ and $\Gamma_{BC}^A$ break into
various parts out of which the only nonzero ones are,
\ba
&S_{\mu\nu\lambda}=\frac{1}{2}[\partial_\mu b_{\nu\lambda}
+\partial_{\nu}b_{\lambda\mu}+\partial_\lambda b_{\mu\nu}];&{}
\label{s0}\\
&S_{\mu ij}=\frac{1}{2}\partial_\mu B_{ij};&{}\label{s1}\\
&\Gamma^{\lambda}_{\mu\nu}
=\frac{1}{2}g^{\lambda\sigma}[\partial_\mu g_{\sigma\nu}
+\partial_\nu g_{\sigma\mu} - \partial_\sigma g_{\mu\nu}];&{}
\label{gamma0}\\
&\Gamma^{\mu}_{ij}
= - \frac{1}{2}g^{\mu\nu}\partial_\nu G_{ij};&{}\label{gamma1}\\
&\Gamma^{i}_{\mu j}
=\frac{1}{2}G^{il}\partial_\mu G_{lj}&{}.\label{gamma2}
\ea

For this form of the background fields the total action can be
written in a form which shows the internal and target space
coordinate indices explicitely. This is necessary to find the
duality invariant form. We start with the four fermion term  in
the following subsection.

\begin{flushleft}
{\sc 4.1 four fermion terms}
\end{flushleft}

The four fermion terms of the action $S_{ff}$ are given by
\be
S_{ff} \equiv \frac{1}{16\pi}\int d^2\sigma {\cal R}_{ABCD}
[{\bar\psi}^A(1+\gamma^5)\psi^C][{\bar\psi}^B(1+\gamma^5)\psi^D].
\label{ff}
\ee
In general ${\cal R}_{ABCD}$ could give sixteen different terms
because any of the four indices could take two values $\mu$ or $i$.
It turns out that for our particular form of the background fields
${\cal R}_{ABCD}$ vanishes unless the number of internal (or
space-time) indices is even. This means we get only the following
eight different terms
\be
S_{ff}=\Sigma_{i=1}^{8} S_{ff}^{(i)}
\ee
where $S_{ff}^{(i)}$'s are once again given in Appendix A.
We now consider each term of this action. First term is already
duality invariant. We take the last term containing only internal
fermions:
\be
S_{ff}^{(8)} \equiv \frac{1}{16\pi}\int d^2\sigma \{
{\cal R}_{ijkl}
[{\bar\psi}^i(1+\gamma^5)\psi^k]
[{\bar\psi}^j(1+\gamma^5)\psi^l]
\}.
\ee
This can be written as
\be
S_{ff}^{(8)}=\Sigma_{i=1}^{8}{\tilde S}_i
\ee
where explicit form of various $\tilde S$'s are given in Appendix A.
Using the identity
\ba
&[\bar{\psi}^i(1+\gamma^5)\psi^k][\bar{\psi}^j(1+\gamma^5)\psi^l]
\nonumber\\
&=-[\bar{\psi}^i(1+\gamma^5)\psi^l]
[\bar{\psi}^j(1+\gamma^5)\psi^k]
\label{fierz0}
\ea
we find that
\be
{\tilde S}_1={\tilde S}_2,
{\tilde S}_3={\tilde S}_4,
{\tilde S}_5={\tilde S}_8,
{\tilde S}_7={\tilde S}_6.
\ee
Therefore we get
\be
S_{ff}^{(8)}=2({\tilde S}_1+{\tilde S}_3+{\tilde S}_5+{\tilde S}_7.)
\ee

We now generalize eq.(\ref{tildepsi}) for curved backgrounds,
\be
\tilde{\psi}_i=G_{ij}\gamma^5\psi^j+B_{ij}\psi^j\label{tildepsi3}.
\ee
Also
\be
\bar{\tilde{\psi}}_i=-G_{ij}\bar{\psi^j}\gamma^5
+B_{ij}\bar{\psi^j}\label{tildepsi4}.
\ee
Then after using the Fierz identities
\be
\bar{\psi}\chi =\bar{\chi}\psi, \;\;\;\;
\bar{\psi}\gamma^5\chi=-\bar{\chi}\gamma^5\psi,\label{fierz}
\ee
we get,
\be
{\tilde S}_1+{\tilde S}_5=
\frac{1}{64\pi}\int d^2\sigma \{
\partial_{\mu}G_{ik}({\bar\psi}^i\psi^k)
[L_{ab}{\bar\psi}^a\gamma^5\partial^\mu \psi_b].\label{1+5.2}
\}
\ee
Similarly
\be
{\tilde S}_3+{\tilde S}_7=
\frac{1}{64\pi}\int d^2\sigma \{
\partial_{\mu}B_{ik}({\bar\psi}^i\gamma^5\psi^k)
[L_{ab}{\bar\psi}^a\gamma^5\partial^\mu \psi_b].\label{3+6}
\}
\ee
Adding (\ref{1+5.2}) and (\ref{3+6}) and using the procedure
we adopted to obtain these equations we get the following
completly duality invariant expression
\be
{\tilde S}_1+{\tilde S}_3+{\tilde S}_5+{\tilde S}_7=
\frac{1}{64\pi}\int d^2\sigma
[L_{ab}{\bar\psi}^a\gamma^5\partial^\mu \psi_b]
[L_{cd}{\bar\psi}^c\gamma^5\partial^\mu \psi_d].\label{1356}
\ee

Thus the total contribution of $S_{ff}^{(8)}$ is
\be
S_{ff}^{(8)}=\frac{1}{32\pi}\int d^2 \sigma\{
[L_{ab}\bar{\psi}^a\gamma^5\partial_\mu \psi^b]
[L_{cd}\bar{\psi}^c\gamma^5\partial^\mu \psi^d]\}.
\label{8.0}
\ee

We now return to other terms in the action $S_{ff}$. Using relations
similar to eq.(\ref{fierz0}) we combine the following terms
\ba
&S_{ff}^{(3)}+ S_{ff}^{(4)}+ S_{ff}^{(6)}+ S_{ff}^{(7)}\nonumber\\
&=\frac{1}{16\pi}\int d^2\sigma({\cal R}_{\mu i\nu j}-
{\cal R}_{\mu ij\nu}+{\cal R}_{i\mu j\nu}-{\cal R}_{i\mu\nu j})
\nonumber\\
&[{\bar\psi}^\mu(1+\gamma^5)\psi^\nu][{\bar\psi}^i(1+\gamma^5)\psi^j]
\label{sff.1}
\ea
The background dependent coefficients can be computed using
the expressions (\ref{s0})-(\ref{gamma2}) and (\ref{data0}).
These are given in eqs.(\ref{data0}) of Appendix A.
Using these it is straightforward to show that
\ba
&S_{ff}^{(3)}+ S_{ff}^{(4)}+ S_{ff}^{(6)}+ S_{ff}^{(7)}\nonumber\\
&=\frac{1}{16\pi}\int d^2 \sigma\{
[\bar{\psi}^\mu (1+\gamma^5) \psi^\nu]
[\partial_\nu (LML)_{ab}\bar{\psi}^a
(1+\gamma^5)\partial^\mu \psi^b]\nonumber\\
&+\frac{1}{2}\Gamma^\alpha_{\mu\nu}
[\bar{\psi}^\mu \psi^\nu]
[L_{ab}\bar{\psi}^a\gamma^5\partial_\alpha\psi^b]\nonumber\\
&-\frac{1}{2}{S^\alpha}_{\mu\nu}
[\bar{\psi}^\mu\gamma^5\psi^\nu]
[L_{ab}\bar{\psi}^a\gamma^5\partial_\alpha\psi^b]\nonumber\\
&-\frac{1}{2}[\bar{\psi}^\mu(1+\gamma^5)\psi^\nu]
[L_{ab}\bar{\psi}^a\gamma^5\partial_\mu\partial_\nu\psi^b]\}.
\label{3467}
\ea

Finally we have to condider the remaining two four fermion terms
\ba
&S_{ff}^{(2)}+ S_{ff}^{(5)}\nonumber\\
&=-\frac{1}{16\pi}\int d^2\sigma\{{\cal R}_{\mu\nu ij}
[{\bar\psi}^\mu(1+\gamma^5)\psi^i][{\bar\psi}^\nu(1+\gamma^5)\psi^j]
\nonumber\\
&+{\cal R}_{ij\mu\nu}
[{\bar\psi}^i(1+\gamma^5)\psi^\mu][{\bar\psi}^j(1+\gamma^5)\psi^\nu]\}
\label{sff25}
\ea
where $R_{\mu\nu ij}$ and  $R_{ij\mu\nu}$ can be calculated using
(\ref{data1}).  Again repeating the familiar steps we arrive at
\ba
&S_{ff}^{(2)}+ S_{ff}^{(5)}\nonumber\\
&=\frac{1}{32\pi}\int d^2 \sigma
(LML)_{ab}[({\bar\psi}^\mu\gamma^5\partial_\nu\psi^a)
({\bar\psi}^\nu\gamma^5\partial_\mu\psi^b)
-({\bar\psi}^\mu\gamma^5\partial_\mu\psi^a)
({\bar\psi}^\nu\gamma^5\partial_\nu\psi^b)\nonumber\\
&+({\bar\psi}^\mu\partial_\nu\psi^a)
({\bar\psi}^\nu\partial_\mu\psi^b)
-({\bar\psi}^\mu\partial_\mu\psi^a)
({\bar\psi}^\nu\partial_\nu\psi^b)].
\ea

We have thus shown that all the four fermion terms can be written
in duality invariant form.

\begin{flushleft}
{\sc 4.2 kinetic term for fermions}
\end{flushleft}

We now discuss the kinetic term for fermions
\be
S^f_{KE}=
-\frac{i}{2\pi}\int d^2\sigma\{
G_{\mu\nu}{\bar\psi}^\mu(\;\slash\!\!\!\!{\cal D}\psi)^\nu +
G_{ij}{\bar\psi}^i(\;\slash\!\!\!\!{\cal D}\psi)^j
\},\label{kef0}
\ee
where,
\ba
({\cal D}_{\halpha}\psi)^\mu &=\partial_{\halpha}\psi^\mu
+\eta_{\halpha\hbeta}\Gamma^\mu_{\lambda\nu}\partial^{\hbeta}
X^{\lambda}\psi^\nu
+\eta_{\halpha\hbeta}\Gamma^\mu_{ij}\partial^{\hbeta}
y^i \psi^j\nonumber\\
&-\epsilon_{\halpha\hbeta}{S^\mu}_{\nu\lambda}\partial^{\hbeta}
X^\nu\psi^\lambda
-\epsilon_{\halpha\hbeta}S^\mu_{ij}\partial^{\hbeta}y^i\psi^j
\ea
and
\ba
({\cal D}_{\halpha}\psi)^i &=\partial_{\halpha}\psi^i
+\eta_{\halpha\hbeta}\Gamma^i_{\mu j}\partial^{\hbeta}
X^{\mu}\psi^j
+\eta_{\halpha\hbeta}\Gamma^i_{\mu j}\partial^{\hbeta}
y^j\psi^\mu \nonumber\\
&-\epsilon_{\halpha\hbeta}G^{ij}S_{jl\mu}\partial^{\hbeta}
y^l \psi^\mu
-\epsilon_{\halpha\hbeta}G^{ij}S_{j\mu l}\partial^{\hbeta}
X^\mu \psi^l.
\ea
We write (\ref{kef0}) as
\be
S^f_{KE}= S^{f0}_{KE}+S^{f1}_{KE}+S^{f2}_{KE}+S^{f3}_{KE}+S^{f4}_{KE}
\label{kef00}
\ee
such that $S^{f0}_{KE}$ is the part unaffected by duality transformations.
The other terms are written as follows.
\ba
S^{f1}_{KE}= -\frac{i}{2\pi}\int d^2\sigma\{
G_{\mu\nu}{\bar\psi}^\mu \gamma^{\halpha}(\eta_{\halpha\hbeta}
\Gamma^\nu_{ij}\partial^{\hbeta}y^i\psi^j -
\epsilon_{\halpha\hbeta}G^{\nu\lambda}S_{\lambda
ij}\partial^{\hbeta}y^i\psi^j)\}\\
\label{kef1}
S^{f2}_{KE}= -\frac{i}{2\pi}\int d^2\sigma\{
G_{ij}{\bar\psi}^i\gamma^{\halpha}(\eta_{\halpha\hbeta}
\Gamma^j_{\mu k}\partial^{\hbeta}X^{\mu}\psi^k
-\epsilon_{\halpha\hbeta}G^{jk}S_{kl\mu}\partial^{\hbeta}
X^\mu \psi^l)\}\\
\label{kef2}
S^{f3}_{KE}= -\frac{i}{2\pi}\int d^2\sigma\{
\psi^i\gamma^{\halpha}G_{ij}(\eta_{\halpha\hbeta}\Gamma^j_{\mu l}
\partial^{\hbeta}y^l
-\epsilon_{\halpha\hbeta}G^{jl}S_{lk\mu}\partial^{\hbeta}y^k)
\psi^\mu\}\\
\label{kef3}
S^{f4}_{KE}= -\frac{i}{2\pi}\int d^2\sigma\{
G_{ij}{\bar\psi}^i\gamma^{\halpha}\partial_{\halpha}\psi^j
\}\label{kef4}
\ea

We start with $S^{f1}_{KE}$ for which we find the following
duality invariant form
\be
S^{f1}_{KE}= -\frac{i}{4\pi}\int d^2\sigma\{
{\bar\psi}^\mu\gamma^5
\gamma^{\halpha}\eta_{\halpha\hbeta}L_{ab}\partial^{\hbeta}y^a
\partial_\mu\psi^b\}.\label{ke1}
\ee
Next we take $S^{f2}_{KE}$. Using the Fierz identities one finds
that first term of eq.(\ref{kef2}) vanishes. We combine its
second term with $S^{f4}_{KE}$ to write the following
duality invariant expression
\be
S^{f2}_{KE}+S^{f4}_{KE}= -\frac{i}{4\pi}\int d^2\sigma\{
L_{ab}{\bar\psi}^a\gamma^5
\gamma^{\halpha}\partial_{\halpha}\psi^b
\}.\label{kef24}
\ee
Another equivalent duality invariant form for these terms is
\be
S^{f2}_{KE}+S^{f4}_{KE}= -\frac{i}{4\pi}\int d^2\sigma\{
(LML)_{ab}{\bar\psi}^a\gamma^{\halpha}\partial_{\halpha}\psi^b
\}.\label{kef25}
\ee
 Finally the term $S^{f3}_{KE}$ can be put into the following
form
\be
S^{f1}_{KE}= -\frac{i}{4\pi}\int d^2\sigma\{
\eta_{\halpha\hbeta}L_{ab}\partial_\mu
{\bar\psi}^a\partial^{\hbeta}y^b\gamma^5\gamma^{\halpha}\psi^\mu
\}.\label{ke4}
\ee
Thus we have obtained the duality invariant form of the total action.

\begin{flushleft}
{\sc 2.3 supersymmetry transformations}
\end{flushleft}

Finally we discuss some issues related to supersymmetry
transformations. The usual supersymmetric transformations for the
nonlonear sigma models in arbitrary backgrounds are
\ba
&\delta_S X^A = \bar{\epsilon} \psi^A&{}\nonumber\\
&\delta_S\psi^A
= -i\gamma^{\halpha}\partial_{\halpha}X^A\epsilon +
\frac{1}{2}\epsilon(\Gamma^A_{BC}{\bar\psi}^B\psi^C+
S^A\;_{BC}{\bar\psi}^B\gamma^5\psi^C)&{}.\label{tr2}
\ea

These transformations are separated into internal and
external coordinates and for the bosonic part we have, as before
\be
\delta_S X^\mu=\bar{\epsilon} \psi^\mu;\;\;\;
\delta_S y^a=\bar{\epsilon} \psi^a
\ee
and for $\psi^\mu$ we have
\ba
\delta_S\psi^\mu&=-i\gamma^{\halpha}\partial_{\halpha}X^\mu\epsilon +
\frac{1}{2}\epsilon(\Gamma^\mu_{\nu\lambda}{\bar\psi}^\nu\psi^\lambda
+S^\mu\;_{\nu\lambda}{\bar\psi}^\nu\gamma^5\psi^\lambda)\nonumber\\
&-\frac{1}{4}G^{\mu\nu}({\bar\psi}^i\psi^j\partial_\nu G_{ij}
+{\bar\psi}^i\gamma^5\psi^j\partial_\nu B_{ij})
\ea
which can be written in the following duality invariant form
\be
\delta_S\psi^\mu=-i\gamma^{\halpha}\partial_{\halpha}X^\mu\epsilon +
\frac{1}{2}\epsilon(\Gamma^\mu_{\nu\lambda}{\bar\psi}^\nu\psi^\lambda+
S^\mu\;_{\nu\lambda}{\bar\psi}^\nu\gamma^5\psi^\lambda)
-\frac{1}{4}G^{\mu\nu}L_{ab}\psi^a\gamma^5\partial_\nu\psi^b.
\ee

Next we consider the supersymmetry transformations for the
internal fermions. The usual transformation, which follows from
eq.(\ref{tr2}), is
\be
\delta_S\psi^i=-i\gamma^{\halpha}\partial_{\halpha}y^i\epsilon
+\frac{1}{2} G^{ij} \epsilon{\bar\psi}^\mu
\gamma^5\partial_\mu{\tilde{\psi}}_j.\label{old}
\ee
Again the bosonic term in this transformation can be modified
using $\tilde y$ equation of motion (\ref{tyeom}). When
combined with the corresponding transformation for ${\tilde\psi}$
the modified supersymmetry transformation for internal fermions can
be written in the following duality invariant form:
\be
\delta_S\psi^a=-i\gamma^1\partial_1y^a\epsilon
+i\gamma^0{(ML)^a}_b\partial_1y^b\epsilon
+\frac{1}{2}(ML)^{ab}\epsilon{\bar\psi}^\mu
\gamma^5\partial_\mu{\tilde{\psi}}_b.\label{new1}
\ee
\sect{5. Conclusions}

In this article we gave duality invariant formulations of
worldsheet actions for superstrings in flat as well as
curved backgrounds. We found that, in both cases, the actions
have a modified Lorentz invariance and supersymmetry. On-shell
these are equivalent to to the usual symmetries. We encountered
a novel situation in which the supersymmetry is off-shell
closed. We also found the duality invariant forms for the
supercurrent, supersymmetric transformations and the equations
of motion.

 It will be very interesting to discuss the quantization
of these duality invariant actions. In particualar it should
be verified that the critical dimensions remains ten even
though we have introduced new 'auxiliary' fields as in
case of bosonic strings.

	Another problem is to write the duality invariant
action for the more interesting case of heterotic strings
including the fermions. We hope to return to these questions
in near future.

\begin{flushleft}
{\sc ACKNOWLEDGEMENTS}
\end{flushleft}

I am thankful to Alok Kumar for suggesting the problem, discussions,
and reading the manuscript. I also thank Ashoke Sen for numerous
discussions and suggestions.

\newpage
\sect{Appendix A}
\setcounter{equation}{0}
\renewcommand{\theequation}{A.\arabic{equation}}

In this Appendix we collect some of the notations and definitions.
First of all we need the following definitions for the four fermion
term of the supersymmetric nonlinear sigma model:
\be
{\cal R}_{ABCD}=R_{ABCD}+S_{ABCD},\label{scriptr}
\ee
\be
S_{ABCD}=\Sigma_{ABCD}+T_{ABCD},
\ee
\be
R_{ABCD}=g_{AE}[\partial_C\Gamma^E_{BD}-\partial_D\Gamma^E_{BC}
+\Gamma^F_{BD}\Gamma_{FC}^E-\Gamma^F_{BC}\Gamma^E_{FD}],
\ee
\be
\Gamma_{BC}^A=\frac{1}{2}g^{AD}[\partial_B g_{DC}+\partial_C g_{DB}
-\partial_D g_{BC}],
\ee
\be
S_{ABC}=\frac{1}{2}[\partial_Ab_{BC}+\partial_Cb_{AB}+\partial_B
b_{CA}],
\ee
\be
\Sigma_{ABCD}=S_{AEC}{S^E}_{BD}-S_{AED}{S^E}_{BC},
\ee
\be
T_{ABCD}=g_{AE}[S^E_{{BC};D} - S^E_{{BD};C}].\label{t}
\ee
Next when separated into compact and noncompact indices we get
the following eight different terms in the four fermion part
of the action
\ba
S_{ff}^{(1)}=-\frac{1}{16\pi}\int d^2\sigma \{
{\cal R}_{\mu\nu\alpha\beta}
[{\bar\psi}^\mu (1+\gamma^5)\psi^\alpha ][{\bar\psi}^\nu
(1+\gamma^5)\psi^\beta]\},\label{ff1}\\
S_{ff}^{(2)}=-\frac{1}{16\pi}\int d^2\sigma \{
{\cal R}_{\mu\nu ij}
[{\bar\psi}^\mu (1+\gamma^5)\psi^i ][{\bar\psi}^\nu
(1+\gamma^5)\psi^j]\},\label{ff2}\\
S_{ff}^{(3)}=-\frac{1}{16\pi}\int d^2\sigma \{
{\cal R}_{\mu i \nu j}
[{\bar\psi}^\mu (1+\gamma^5)\psi^\nu ][{\bar\psi}^i
(1+\gamma^5)\psi^j]\},\label{ff3}\\
S_{ff}^{(4)}=-\frac{1}{16\pi}\int d^2\sigma \{
{\cal R}_{\mu i j \nu}
[{\bar\psi}^\mu (1+\gamma^5)\psi^j ][{\bar\psi}^i
(1+\gamma^5)\psi^\nu]\},\label{ff4}\\
S_{ff}^{(5)}=-\frac{1}{16\pi}\int d^2\sigma \{
{\cal R}_{ij\mu\nu}
[{\bar\psi}^i (1+\gamma^5)\psi^\mu ][{\bar\psi}^j
(1+\gamma^5)\psi^\nu]\},\label{ff5}\\
S_{ff}^{(6)}=-\frac{1}{16\pi}\int d^2\sigma \{
{\cal R}_{i\mu j\nu}
[{\bar\psi}^i (1+\gamma^5)\psi^j ][{\bar\psi}^\mu
(1+\gamma^5)\psi^\nu] \},\label{ff6}\\
S_{ff}^{(7)}=-\frac{1}{16\pi}\int d^2\sigma \{
{\cal R}_{i\mu\nu j}
[{\bar\psi}^i (1+\gamma^5)\psi^\nu ][{\bar\psi}^\mu
(1+\gamma^5)\psi^j]\},\label{ff7}\\
S_{ff}^{(8)}=-\frac{1}{16\pi}\int d^2\sigma \{
{\cal R}_{ijkl}
[{\bar\psi}^i(1+\gamma^5)\psi^k]
[{\bar\psi}^j(1+\gamma^5)\psi^l]
\}.\label{ff8}
\ea

$S_{ff}^{(8)}$ has the following parts
\ba
{\tilde S}_1= \frac{1}{16\pi}\int d^2\sigma \{
G_{ip}\Gamma^{\mu}_{jl}\Gamma^{p}_{\mu k}
[{\bar\psi}^i(1+\gamma^5)\psi^k]
[{\bar\psi}^j(1+\gamma^5)\psi^l]\},\\
{\tilde S}_2= -\frac{1}{16\pi}\int d^2\sigma \{
G_{ip}\Gamma_{jk}^{\mu}\Gamma^{p}_{\mu l}
[{\bar\psi}^i(1+\gamma^5)\psi^k]
[{\bar\psi}^j(1+\gamma^5)\psi^l]\},\\
{\tilde S}_3= -\frac{1}{16\pi}\int d^2\sigma \{
S_{i\mu k}S^{\mu}\;_{jl}
[{\bar\psi}^i(1+\gamma^5)\psi^k]
[{\bar\psi}^j(1+\gamma^5)\psi^l]\},\\
{\tilde S}_4= \frac{1}{16\pi}\int d^2\sigma \{
S_{i\mu l}S^{\mu}\;_{jk}
[{\bar\psi}^i(1+\gamma^5)\psi^k]
[{\bar\psi}^j(1+\gamma^5)\psi^l]\},\\
{\tilde S}_5= -\frac{1}{16\pi}\int d^2\sigma \{
G_{ip}\Gamma_{k\mu}^pS^{\mu}\;_{jl}
[{\bar\psi}^i(1+\gamma^5)\psi^k]
[{\bar\psi}^j(1+\gamma^5)\psi^l]\},\\
{\tilde S}_6= -\frac{1}{16\pi}\int d^2\sigma \{
G_{ip}\Gamma_{lj}^{\mu}S^{p}\;_{\mu k}
[{\bar\psi}^i(1+\gamma^5)\psi^k]
[{\bar\psi}^j(1+\gamma^5)\psi^l]\},\\
{\tilde S}_7= \frac{1}{16\pi}\int d^2\sigma \{
G_{ip}\Gamma_{kj}^{\mu}S^p\;_{\mu l}
[{\bar\psi}^i(1+\gamma^5)\psi^k]
[{\bar\psi}^j(1+\gamma^5)\psi^l]\},\\
{\tilde S}_8= \frac{1}{16\pi}\int d^2\sigma \{
G_{ip}\Gamma_{l\mu}^pS^{\mu}\;_{jk}
[{\bar\psi}^i(1+\gamma^5)\psi^k]
[{\bar\psi}^j(1+\gamma^5)\psi^l]
\}.
\ea
We need the following expressions to write (\ref{sff.1})
in duality invariant form
\ba
&R_{\mu i\nu j}= -R_{\mu ij\nu}=R_{i\mu j\nu}=
-R_{i\mu\nu j}=\nonumber\\
&\frac{1}{2}\Gamma^\alpha_{\mu\nu}\partial_\alpha G_{ij}
-\frac{1}{2}\partial_\mu\partial_\nu G_{ij}
+\frac{1}{4}G^{lm}\partial_\mu G_{mj}\partial_\nu G_{il}\nonumber\\
&\Sigma_{\mu i\nu j}= -\Sigma_{\mu ij\nu}=
\Sigma_{i\mu j\nu}= -\Sigma_{i\mu\nu j}=\nonumber\\
&\frac{1}{4}G^{km}\partial_\mu B_{mj}
\partial_\nu B_{ik} -\frac{1}{2}S^\alpha\;_{\mu\nu}
\partial_{\alpha}B_{ij}\nonumber\\
&T_{\mu i\nu j}= -T_{\mu ij\nu}=
T_{i\mu j\nu}= -T_{i\mu\nu j}=\nonumber\\
&-\frac{1}{2}\partial_\mu \partial_\nu B_{ij}
-\frac{1}{2}S^\alpha\;_{\mu\nu}\partial_\alpha G_{ij}\nonumber\\
&+\frac{1}{2}\Gamma^\alpha\;_{\mu\nu}\partial_\alpha B_{ij}
+\frac{1}{4}G^{lm}(\partial_\mu G_{mj}\partial_\nu B_{il}
+\partial_\nu G_{il}\partial_\mu B_{mj}).
\label{data0}
\ea
To put (\ref{sff25}) in duality invariant form we need
\ba
&R_{\mu\nu ij}=R_{ij\mu\nu}=\frac{1}{4}G^{lm}
(\partial_\nu G_{im}\partial_\mu G_{lj}
-\partial_\mu G_{im}\partial_\nu G_{lj}),\nonumber\\
&\Sigma_{\mu\nu ij}=\Sigma_{ij\mu\nu}=\frac{1}{4}G^{lm}
(\partial_\mu B_{im}\partial_\nu B_{lj}
-\partial_\nu B_{im}\partial_\mu B_{lj}),\nonumber\\
&T_{\mu\nu ij}=-T_{ij\mu\nu}=\frac{1}{4}G^{lm}
(-\partial_\mu G_{im}\partial_\nu B_{lj}
+\partial_\nu G_{im}\partial_\mu B_{lj}
\nonumber\\
&-\partial_\nu B_{im}\partial_\mu G_{lj}
+\partial_\mu B_{im}\partial_\nu G_{lj}).
\label{data1}
\ea

\end{document}